\documentclass[12pt,a4paper]{article}
\usepackage[english]{babel}
\usepackage{amsmath,amssymb,amsfonts}
\usepackage{bm,latexsym,euscript,textcomp}
\usepackage{graphicx,epsfig,cite}

\newcommand{\beq}{\begin{equation}}
\newcommand{\eeq}{\end{equation}}
\newcommand{\pt}{\partial}

\begin{document}

\title{Condensation of water vapor in the gravitational field}

\author{V. G. Gorshkov, A. M. Makarieva, A. V. Nefiodov\thanks{E-mail: anef@thd.pnpi.spb.ru}}

\date{B. P. Konstantinov Petersburg Nuclear Physics Institute, \\
188300, Gatchina, Leningrad district, Russia}

\maketitle

\begin{abstract}
Physical peculiarities of water vapor condensation under conditions of hydrostatic equilibrium are considered. The power of stationary dynamic air fluxes and the vertical temperature distribution caused by condensation on large horizontal scales are estimated.
\end{abstract}


\section{Introduction}
\label{sec1}
Upon a temperature decrease, saturated vapor undergoes condensation, which results in a drop of the gas pressure and brings about a dynamic gas flow from the area of high temperature to the area of low temperature. This phenomenon is widely used in technical applications like, for example, the heat pipe \cite{gro64,xie08}. In most empirical and theoretical investigations of this effect the condensation occurs on a solid surface that serves as a rigid boundary for the dynamic flow, while the temperature of the surface where condensation occurs is set by external conditions \cite{sha90,son00,ryk09}. The characteristic linear scale of such problems in laboratory does not exceed several meters.

Condensation-related dynamic gas flows that arise at much larger linear scales in the gravitational field of the Earth have a number of peculiar features. The drop of temperature that induces condensation occurs during the vertical upward motion of an air volume as a consequence of the increase in the potential energy of the ascending air at the expense of a decrease in its internal energy. Therefore, for condensation to occur it is not necessary to maintain artificially a temperature gradient that is limited by the intensity of heat removal to an external environment. Furthermore, condensation occurs within a moving air volume rather than on a macroscopic rigid surface that bounds the dynamic flow.

Air is maintained at the planetary surface by the gravitational field of the Earth; its vertical distribution is governed by the condition of hydrostatic equilibrium. Water vapor condensation disturbs the hydrostatic distribution of air and leads to the appearance of the upward-directed pressure gradient forces that are not compensated by the gravitational field. The vertical air flow is stopped by the gravitational field that tends to restore the hydrostatic equilibrium. As a result, in an open (horizontally unbounded) space, if condensation takes place, there arises a horizontal air flow and a horizontal pressure gradient force that is directed inward the condensation area. In this paper the peculiarities of water vapor condensation under condition of hydrostatic equilibrium are considered that have a number of applications \cite{pl11a,pl11b}.

\section{The continuity equation with an account of condensation}
\label{sec2}

In the stationary case the continuity equation for air has the following form
\begin{gather}
\mathrm{div} N\textbf{v} = {\cal S},   \quad  \textbf{v} = \textbf{u} + \textbf{w} \,  , \label{eq1}\\
\mathrm{div} N_v \textbf{v}= {\cal S}, \quad  \mathrm{div} N_d \textbf{v} = 0, \quad N=N_d+N_v  \,  ,  \label{eq2}
\end{gather}
where $N$, $N_d$, è $N_v$ [mol/m$^3$] are the molar densities of moist air, dry air and saturated water vapor,  respectively, ${\cal S}$ [mol/m$^3 \cdot$ s] is the density of condensation rate, $\textbf{u}$ and $\textbf{w}$ are the air velocities in the horizontal and vertical directions, respectively.
We choose the axes such that the $x$ axis is directed along the horizontal velocity $\textbf{u}$,
while the $z$ axis is parallel to the vertical velocity $\textbf{w}$. Saturated molar density $N_v$ of water vapor depends, in accordance with the Clausius-Clapeyron law, on absolute temperature $T$ only. Assuming the air to be isothermal in the horizontal plane at any $z$ we have:
\begin{equation}\label{eq3}
\frac{\partial T}{\partial x}= \frac{\partial N_v}{\partial  x}=0\,  , \quad N_v=N_v(T)\,  , \quad
T=T(z) \,  .
\end{equation}
Let us introduce dimensionless variables $\gamma$ and $\gamma_d$ according to the following definitions:
\begin{equation}\label{eq4}
\gamma\equiv \frac{N_v}{N} \,  , \quad  \gamma_d\equiv \frac{N_v}{N_d}\,  , \quad
\gamma\equiv \frac{\gamma_d}{1+\gamma_d} \,  , \quad \gamma_d\equiv \frac{\gamma}{1-\gamma}
\,  .
\end{equation}
In the terrestrial atmosphere the quantities $\gamma$ and $\gamma_d$ are small and do not exceed $0.1$. Multiplying the second equation in \eqref{eq2}, that contains $N_d$, by $\gamma_d$ and subtracting from it the first equation in \eqref{eq2}, that contains $N_v$, we find that the continuity equations \eqref{eq2} take the following form:
\begin{equation}\label{eq5}
u \frac{\partial N}{\partial  x}= u \frac{\partial N_d}{\partial  x}= \left( S_d - {\cal S}\right) \frac{1}{\gamma_d} \,  , \quad S_d \equiv w \left( \frac{\partial N_v}{\partial z} - \gamma_d \frac{\partial N_d}{\partial z}\right) \equiv w N_d \frac{\partial \gamma_d}{\partial  z} \,  .
\end{equation}
Let us also introduce the quantity $S$ that arises from $S_d$ after $N_d$ is replaced by $N$:
\begin{equation}\label{eq6}
S\equiv w \left( \frac{\partial N_v}{\partial z} - \gamma \frac{\partial N}{\partial z}\right) \equiv w N \frac{\partial \gamma}{\partial  z} \,  , \quad S_d \equiv \frac{S}{1-\gamma}  \,  , \quad
\left( S_d - S\right)\frac{1}{\gamma_d} \equiv  S  \,  .
\end{equation}
The identities written behind the definition of $S$ can be easily checked.

Using the ideal gas equation of state
\begin{equation}\label{eq7}
p= N RT \,  , \quad p_d = N_d RT \,  , \quad p_v= N_v RT
\,  ,
\end{equation}
where $R=8.31$ J/mol$\cdot$K is the universal molar gas constant, $p$, $p_d$ and $p_v$ are the pressures of moist air, dry air components and water vapor, respectively, one can rewrite relations \eqref{eq5} in terms of the corresponding pressures:
\begin{gather}
u \frac{\partial p}{\partial  x}= u \frac{\partial p_d}{\partial  x}= \left( s_d - \sigma\right) \frac{1}{\gamma_d} \,  , \quad  s_d \equiv \frac{s}{1-\gamma} \,  , \quad
\sigma \equiv{\cal S}RT\,   , \label{eq8}\\
s_d \equiv w \left( \frac{\partial p_v}{\partial z} - \gamma_d \frac{\partial p_d}{\partial z}\right) \equiv w p_d \frac{\partial \gamma_d}{\partial  z}  \,  , \quad
s \equiv w \left( \frac{\partial p_v}{\partial z} - \gamma \frac{\partial p}{\partial z}\right) \equiv w p \frac{\partial \gamma}{\partial  z} \,  .   \label{eq9}
\end{gather}
Partial derivatives $\partial  T/\partial  z$ cancel from \eqref{eq7} due to the universality of the gas constant and do not appear in equation \eqref{eq8}. Continuity equations \eqref{eq1} and \eqref{eq2} written in the form of \eqref{eq5} and \eqref{eq8} do not, in the view of \eqref{eq7}, contain mass densities $\rho_i = M_iN_i$ and molar masses $M_i$ $(i = v, d)$.

The quantities $\sigma$, $s_d$ and $s$ [W/m$^3$] have the meaning of the dynamic power densities of condensation. According to \eqref{eq8}, at any $\sigma$ that differs from $s_d$ by a small magnitude of the order of $\gamma_d$, there appears a horizontal pressure gradient. In particular, at $\sigma = s$ \eqref{eq9} we have $u \partial p/\partial x  = s$. Neglecting terms of the order of $\gamma_d$ in numerical models describing the moist air circulation (see, e.g., \cite{gill,smi97}) is erroneous. The horizontal pressure gradient disappears in the case of an exact equality $\sigma = s_d$ only.

\section{The physics of the condition of hydrostatic equilibrium}
\label{sec3}

Let us consider the condition of hydrostatic equilibrium and its physical meaning in greater detail.
The equation of states \eqref{eq7} can be re-written as follows:
\begin{gather}
p=\rho g h_g\,  , \quad p_d= \rho_d g h_d \,  , \quad p_v= \rho_v g h_v  \,  , \quad p=p_d+p_v \,  , \label{eq10}\\
h_g \equiv \frac{RT}{Mg} \,  , \quad h_d \equiv \frac{RT}{M_d g} \,  , \quad h_v \equiv \frac{RT}{M_v g} \,  , \quad M = M_d(1-\gamma) + \gamma M_v  \,  , \label{eq11}\\
\rho=\rho_d+ \rho_v \,   , \quad \rho= M N\,   , \quad \rho_d=M_d N_d \,  , \quad  \rho_v=M_v N_v \,  ,  \label{eq12}
\end{gather}
where $\gamma$ is defined in \eqref{eq4}, $\rho$, $\rho_d$,  $\rho_v$ are the mass densities of the gases, $M_v = 18$ g/mol, $M_d = 29$ g/mol are the molar masses of the gases, $h_g$, $h_d$ and  $h_v$ are the heights of the uniformly dense atmospheres consisting of moist air, its dry component and water vapor, respectively. In the form of \eqref{eq10} it is immediately clear that the gas pressures represent the densities of potential energy in the gravitational field of the Earth. The conditions of hydrostatic equilibrium are:
\begin{equation}\label{eq13}
- \frac{\partial  p}{\partial  z}= \rho g \equiv \frac{p}{h_g}\,  , \quad
- \frac{\partial  p_d}{\partial  z}= \rho_d g \equiv \frac{p_d}{h_d}\,  , \quad
- \frac{\partial  p_v}{\partial  z}= \rho_v g \equiv \frac{p_v}{h_v}\,  .
\end{equation}

In motionless air in the absence of condensation all the three equalities in \eqref{eq13} are fulfilled. As the moist air ascends, all the gas components move at one and the same velocity $w$. Therefore, if there is no condensation (when it is the unsaturated water vapor that moves), hydrostatic equilibrium is established in such a manner that at all heights the mean molar mass of the air mixture is equal to its value at the surface, i.e., that $h_g$ does not depend on $z$. With the onset of condensation the last equality in \eqref{eq13} for $p_v$ is violated: the vertical distribution of water vapor is compressed \cite{hess07} (see also Section 5, formulae \eqref{eq22}, \eqref{eq28}):
\beq\label{hc}
-\frac{\pt p_v}{\pt z} = \frac{p_v}{h_c},\quad h_c < h_g.
\eeq
There appears an upward force $f_c$ that acts on the remaining moist air. The corresponding power $wf_c$ is equal to (see \eqref{eq9})
\beq\label{wfc}
wf_c = w\left(\frac{p_v}{h_c} - \frac{p_v}{h_g}\right) = -w\left(\frac{\pt p_v}{\pt z} - \frac{p_v}{p} \frac{\pt p}{\pt z}\right)\equiv -s,\,\,\,h_c < h_g.
\eeq
In an open space the disturbed hydrostatic equilibrium is re-established at the expense of a horizontal air inflow inward the condensation area, i.e. there arise a horizontal pressure gradient and a horizontal force, while the vertical force disappears. It follows from the energy conservation law that the power of circulation that in hydrostatic equilibrium is determined by the horizontal pressure gradient and horizontal velocity only, $-u\pt p/\pt x$, is equal to power $wf_c$, which corresponds to \eqref{eq8} at $\sigma = s$:
\beq\label{power}
wf_c = -u\frac{\pt p}{\pt x} = -s.
\eeq
Thus in an open space with the condensation present the fulfilment of the condition of hydrostatic equilibrium for moist air as a whole corresponds to the following equalities:
\begin{equation}\label{eq14}
- \frac{\partial  p}{\partial  z}= \frac{p}{h_g} \,  , \quad -\frac{\partial  p_d}{\partial  z} = \frac{p}{h_g} - \frac{p_v}{h_c},\quad
- \frac{\partial  p_v}{\partial  z} =\frac{p_v}{h_c}.
\end{equation}
In a horizontally bounded space where $\pt p/\pt x = 0$, the compensation of deviation from hydrostatic equilibrium caused by the condensation and the disappearance of the vertical force $f_c$ are not possible at $\gamma < 1$ (see \eqref{eq36} in Section 5).

Immediately upon condensation of the water vapor the cumulative density of the gas $\rho_v$ and condensed $\rho_l$ phases remains unchanged: $\rho_v + \rho_l  = const$. However, Eq.~\eqref{eq7} depends on the total number of gas particles (or, in the established units of measurements, the total number of moles) per unit volume independent of their mass and size. When the liquid or solid phase and droplets are formed, the original number of vapor particles decreases by more than millions of times. Partial pressure of the arising Brownian particles of the condensed phase is decreased by the same factor. Therefore, the first equation in \eqref{eq14} depends on the number of particles in the gas phase only (to the relative accuracy of the order of $10^{-6}$), irrespective of whether the condensed particles remain in the unit volume or fall out from it under the action of gravity. Thus in \eqref{eq14} we have
$$\rho= \rho_d +\rho_v \neq \rho_d + \rho_v + \rho_l.$$
Using the last equality that contains $\rho_l$, as is commonly done in numerical models \cite{bry09}, is erroneous \cite{pel11}.

The friction force for the ascending air flow due to rain drops falling at constant velocity, which is dictated by the Stokes law \cite{land}, is determined by the volume density of the number of drops and their radius, i.e. it is governed by independent physical principles. Therefore, in the general case, this friction force cannot determine the presence or absence of hydrostatic equilibrium, which is satisfied to a very high accuracy in large-scale circulation patterns, with taking into account of the falling drops \cite{pl09a,wat10}. At the same time, the rain friction force along with the force of surface friction that is proportional to the weight of the atmospheric column is a force that impedes horizontal acceleration of the air flow in large-scale circulation patterns \cite{pl09a,wat10}.

\section{Condensation fluxes in open space in the gravitational field of the Earth}
\label{sec4}

In an open space, stationarity of the condensation process in the absence of a horizontal temperature gradient can be maintained by continuous evaporation from the liquid surface of the Earth that compensates condensation at all heights. Evaporation with intensity lower than the intensity of condensation in the entire atmospheric column ensures that the $x$-independent height $z$, where the relative humidity reaches unity, is constant with time. In the case when the condensation power significantly exceeds the power of evaporation, the stationarity of the former quantity can only be maintained if the wind pattern moves towards areas with saturated water vapor that occurs in cyclones, hurricanes and tornadoes \cite{pl09b,pl11a,pl11b}.

Condensation power on retention of the hydrostatic distribution of moist air \eqref{eq14} is determined by the value of $\sigma = s$ \eqref{eq9}. Condensation power $s$ \eqref{eq9} has a simple physical meaning, see \eqref{hc}-\eqref{power}: it is equal to the power of the change of $p_v$ over $z$ as the air ascends with a vertical velocity $w$ minus that of the change of $p_v$, which is not related to condensation -- it is proportional to the relative power of the change of the pre-condensation equilibrium distribution of moist air as a whole. In this case, in agreement with \eqref{eq8}, there arises a power of the horizontal pressure gradient $u \partial p/\partial x$, which compensates the power of the on-going condensation and serves to preserve the vertical hydrostatic distribution of moist air. Equations \eqref{eq8} and \eqref{eq5} take the following form
\begin{gather}
u \frac{\partial  p}{\partial  x}= w \left(\frac{\partial  p_v}{\partial  z} - \gamma   \frac{\partial  p}{\partial  z} \right)\,  , \quad \sigma=s= p w \frac{\partial \gamma}{\partial z}= S RT \, , \label{eq15}\\
u \frac{1}{\gamma}\frac{\partial \gamma}{\partial  x}= -w \frac{\pt \gamma}{\partial  z}\,  , \quad
\frac{1}{p}\frac{\pt p}{\pt x} = -\frac{1}{\gamma}\frac{\pt \gamma}{\pt x}\, , \quad
\frac{\pt p_v}{\pt x}=0\,  , \label{eq16}
\end{gather}
and hereby represent the main dynamic equations that describe the condensation-induced air circulation. Condensation changes the vertical distribution of both water vapor and dry air components, see \eqref{eq14}.

By virtue of the Euler's equations or the Bernoulli integral the appearance of a horizontal pressure gradient results in a horizontal air flow. The vertical force remains compensated by gravity under condition of hydrostatic equilibrium. When the friction forces are small compared to the pressure gradient force the air accelerates in the horizontal plane, an example of which is given by the air masses converging  to the center of condensation area in hurricanes \cite{pl09a,pl11a} and tornadoes \cite{pl11b}. When the magnitude of the horizontal pressure gradient force and that of the friction force coincide, the horizontal inflow of air inward the condensation area is characterized by a constant velocity \cite{hess07,wat10}.

That a horizontal pressure gradient \eqref{eq15} must arise in a large-scale circulation with
the horizontal length scale $L$ exceeding its height $h_g$ by two orders of magnitude is physically obvious:
when height $h_g$ is limited, irrespective of the nature of physical forces that ensure its finite value, the condensation-induced pressure drop can be distributed in the horizontal dimension only. The integral form of the continuity equation \eqref{eq1} for the entire circulation area at $\gamma \ll 1$ is given by $u/L = w/h$. Neglecting friction the energy conservation (Bernoulli equation) corresponds to the following relations $\Delta p_z \sim  \rho w^2/2$, $\Delta p_x \sim  \rho u^2/2$. At $L \sim 3\times 10^3$ km, $h \sim 10$ km we have
$$\frac{\Delta p_z }{\Delta p_x} = \frac{w^2}{u^2} = \frac{h^2}{ L^2} \simeq 10^{-5} \ll 1.$$
The existence of a single dimensional scale for the total pressure drop in a circulation, $\Delta p_z \sim p_v \sim 1$~kPa, that does not depend on the linear size of the circulation, is supported by observations \cite{hol04}: hurricanes, squalls and tornadoes whose linear sizes differ by hundreds and thousands of times are characterized by similar values of the total pressure drop.

Condensation in a horizontally bounded volume can be envisioned in the form of moist air rising in a vertical tube that embraces the entire condensation area. If the distribution of the dry air component remains unchanged, the condensation power is determined by the value of $\sigma = s_d = s /(1- \gamma)$, see~\eqref{eq8}, by analogy to how when the vertical distribution of moist air as a whole is preserved we have $\sigma = s$. Vertical gradients of moist air and water vapor (the first and third equalities in Eq. \eqref{eq13}) deviate from hydrostatic equilibrium. In the absence of hydrostatic equilibrium the condensation brings about an upward pressure gradient force which accelerates the ascending air up to hurricane velocities \cite{pl09b,pl11b}.

\section{The vertical gradient of air temperature at adiabatic condensation}
\label{sec5}

In a horizontally isothermal atmosphere the condensation occurs owing to the vertical adiabatic ascent of a volume $\widetilde{V}$ of moist air. The first law of thermodynamics, with an account of condensation, written for an air volume $\widetilde{V}=\widetilde{N} V$ that contains $\widetilde{N}$ air moles, where $V = N^{-1}$ is molar volume, $N$ is molar density, has the following form for adiabatically ($dQ = 0$) ascending air:
\begin{equation}\label{eq17}
c_V d(\widetilde{N}T)+ p d(\widetilde{N}V)+ ({\cal L}_0- c_lT) d \widetilde{N}_v = 0 \,  ,
\quad \widetilde{N}=\widetilde{N}_d+ \widetilde{N}_v  \,  .
\end{equation}
Here $\widetilde{N}_d$  and  $\widetilde{N}_v$  are the numbers of moles of the dry air component and water vapor, respectively, $Q$, $c_V$ è $c_l$  are the molar heat, heat capacity of air at constant volume and of liquid water, respectively; ${\cal L}_0$ is the latent heat of breaking the intermolecular
bonds in the liquid, the term $-c_lT$ takes into account heat losses from the gas phase that are spent to warm the newly formed drops of liquid \cite{lev}; $p$ is gas pressure,  $T$ is absolute temperature, $c_V\widetilde{N}T$  is the internal energy of $\widetilde{N}$ moles of gas. During condensation in volume $\widetilde{V}$ it is only the number of the moles of water vapor that changes, while the number of moles of the dry air component does not. Expanding the differentials of products in \eqref{eq17}, changing from $dV$ to $dp$ with use of \eqref{eq7} and dividing both parts of \eqref{eq17} by $\widetilde{N}$ we obtain:
\begin{equation}\label{eq18}
c_p dT - Vdp + {\cal L}\frac{d\widetilde{N}_v}{\widetilde{N}}=0 \, , \quad
{\cal L} = {\cal L}_0 + (c_p-c_l)T = {\cal L}(T_0) + (c_p - c_l)(T-T_0)\,  ,
\end{equation}
where ${\cal L}$ is latent heat of vaporization at temperature $T$, $T_0$ is an arbitrary initial value of temperature \cite{lev}. Dividing the left-hand part of \eqref{eq18} by $c_pT$ we have:
\begin{equation}\label{eq19}
\frac{d T}{T} - \mu \frac{dp}{p} + \mu \xi \frac{d \widetilde{N}_v}{\widetilde{N}}=0 \,  ,\quad
\mu  \equiv \frac{R}{c_p}\, , \quad \xi\equiv  \frac{{\cal L}}{RT} \,  , \quad c_p=c_V + R \,  .
\end{equation}
With help of definition \eqref{eq4} we obtain the following relationship between $d\widetilde{N}_v/\widetilde{N}$ and $d\gamma$:
\begin{equation}\label{eq20}
\frac{d\widetilde{N}_v}{\widetilde{N}}=d\left(\frac{\widetilde{N}_v}{\widetilde{N}}\right)+\gamma \frac{d\widetilde{N}}{\widetilde{N}} \, , \quad
\frac{d\widetilde{N}_v}{\widetilde{N}} =\frac{d\gamma}{1-\gamma} \, , \quad  d\widetilde{N} = d\widetilde{N}_v\,  .
\end{equation}
As a result, \eqref{eq19} takes the form:
\begin{equation}\label{eq21}
\frac{d T}{T} - \mu \frac{dp}{p} + \mu \xi \frac{d \gamma}{1-\gamma} = 0\,  .
\end{equation}

Using the Clausius-Clapeyron equation
\begin{equation}\label{eq22}
\frac{dp_v}{p_v} = \xi \frac{dT}{T}
\end{equation}
and the definition of variable $\gamma$ \eqref{eq4} we have:
\begin{equation}\label{eq23}
\gamma\equiv \frac{\widetilde{N}_v}{\widetilde{N}}= \frac{p_v}{p}\,  , \quad
\frac{d\gamma}{\gamma} = \frac{dp_v}{p_v} - \frac{dp}{p} = \xi \frac{dT}{T} - \frac{dp}{p}\,  .
\end{equation}
With help of formulas \eqref{eq21} and \eqref{eq23} we arrive at the following pairwise relations between
three relative total differentials of the variables $p$, $T$ and $\gamma$:
\begin{gather}
\frac{d\gamma}{\gamma} = \frac{dT}{T}\Phi_{\gamma T}\, , \quad \Phi_{\gamma T} \equiv \frac{\mu\xi-1}{\mu(1+
\gamma_d\xi)} \,  , \label{eq24}\\
\frac{d\gamma}{\gamma} = \frac{dp}{p}\Phi_{\gamma p} \, , \quad  \Phi_{\gamma p} \equiv \frac{\mu\xi-1}{1+
\gamma_d\mu\xi^2}\,  , \label{eq25}\\
\frac{dT}{T} = \frac{dp}{p}\Phi_{T p} \, , \quad  \Phi_{T p}  \equiv \mu\frac{1+\gamma_d\xi}{1+\gamma_d\mu\xi^2} = \frac{\Phi_{\gamma p}}{\Phi_{\gamma T}}\,  . \label{eq26}
\end{gather}

In a stationary air flow the variation of all the quantities in \eqref{eq18} with time $t$  has the following form:
\begin{equation}\label{eq27}
\frac{dA}{dt} = w\frac{\partial A}{\partial z} + u \frac{\partial A}{\partial x},
\quad (A = p,\gamma, T)\, .
\end{equation}
From continuity equations \eqref{eq2} and \eqref{eq27} with taking into account \eqref{eq20} we have:
\beq\label{sp}
\frac{1}{\widetilde{N}}\frac{d\widetilde{N}_v}{dt} =\frac{1}{1-\gamma}\frac{d\gamma}{dt} = \frac{\cal S}{N} = \frac{\sigma}{p}.
\eeq
After the power of condensation $\sigma$ is set and the  horizontal isothermality \eqref{eq3} is taken into account, the derivatives with respect to $x$ in \eqref{eq27} become related to the derivatives with respect to $z$ by the continuity equations \eqref{eq1}, \eqref{eq2}, \eqref{eq8}, \eqref{eq9}. Then equation \eqref{eq21} includes the derivatives with respect to $z$ only.

Let us use \eqref{eq23} and the following definitions to introduce characteristic heights as
\begin{equation}\label{eq28}
\frac{1}{h_T}\equiv - \frac{1}{T} \frac{\pt T}{\pt z}\,  , \quad
\frac{1}{h_p}\equiv - \frac{1}{p} \frac{\pt p}{\pt z}\,  , \quad
\frac{1}{h_\gamma}\equiv - \frac{1}{\gamma} \frac{\pt \gamma}
{\pt z} = \frac{\xi}{h_T} - \frac{1}{h_p}\,  .
\end{equation}
Identities \eqref{eq28} define the heights $h_T$, $h_p$ and $h_\gamma$; $h_c \equiv h_T\xi^{-1}$ is the height of the vertical distribution of saturated water vapor according to the Clausius-Clapeyron equation \eqref{eq22} and \eqref{eq3}. Height $h_p$ coincides with $h_g$ \eqref{eq11} if only the condition of hydrostatic equilibrium is fulfilled. The last equality in \eqref{eq28} arises from the Clausius-Clapeyron law and relates the heights $h_\gamma$, $h_T$ and $h_p$ to each other. Equation \eqref{eq21} relates these three heights in the view of \eqref{eq27} and the continuity equations \eqref{eq8}. Thus, two equations, \eqref{eq21} and \eqref{eq22}, relate three unknown heights. To determine the magnitudes of all the three heights it is necessary to set height $h_p$ by some physical condition similar to that considered in Section \ref{sec4}.

When condensation occurs in an open space and the condition of hydrostatic equilibrium of moist air as a whole is preserved, $h_p=h_g$ \eqref{eq14}, according to \eqref{power} and \eqref{eq16} we have:
\begin{equation}\label{eq29}
u\frac{1}{p} \frac{\partial p}{\partial x} = -u\frac{1}{\gamma} \frac{\partial \gamma}{\partial x}=w\frac{\partial \gamma}{\partial z}.
\end{equation}
Note that the condition $-\pt p_d/ \pt z > 0$, see the second equality in \eqref{eq14},
sets a limitation on the value of $\gamma$:
\beq\label{gam}
-\frac{\pt p_d} {\pt z} > 0 \,  , \quad  \gamma < \frac{h_c}{h_d+\varepsilon h_c}\, , \quad \varepsilon \equiv \frac{M_d - M_v}{M_d} = 0.38 .
\eeq

 As a result, for $d\gamma/dt$ \eqref{eq27} and $(d\widetilde{N}_v /dt)/\widetilde{N}$ \eqref{eq20} we obtain:
\begin{equation}\label{eq30}
\frac{d\gamma}{dt}=w(1-\gamma)\frac{\partial \gamma}{\partial z}\, , \quad
\frac{1}{\widetilde{N}}\frac{d{\widetilde N}_v}{dt} = \frac{1}{1-\gamma}\frac{d\gamma}{dt}=w\frac{\partial \gamma}{\partial z} \equiv \frac{s}{p}\,  .
\end{equation}
Using \eqref{eq28}, \eqref{eq30} and omitting the common multiplier $w$ we have from \eqref{eq21}:
\beq\label{eq31}
\frac{1}{h_T} - \mu \frac{1}{h_p} + \gamma\mu (\xi-1) \frac{1}{h_\gamma} = 0\,  .
\eeq
The last equality in \eqref{eq28} together with \eqref{eq31} lead to the following pairwise relations:
\begin{gather}
\frac{1}{h_\gamma}=\frac{1}{h_T}\varphi_{\gamma T} \, , \quad \varphi_{\gamma T} \equiv \frac{\mu\xi-1}{\mu(1+\gamma(\xi-1))} \, , \quad
s=-wp\frac{\gamma}{h_\gamma}\,  , \label{eq32}\\
\frac{1}{h_\gamma}=\frac{1}{h_p}\varphi_{\gamma p} \, , \quad \varphi_{\gamma p} \equiv \frac{\mu\xi-1}{1+\gamma\mu\xi(\xi-1)} \,  , \quad
h_p=h_g \,  , \label{eq33}\\
\frac{1}{h_T}=\frac{1}{h_p}\varphi_{T p}\, , \quad \varphi_{T p} \equiv \mu\frac{1+\gamma(\xi-1)}{1+\gamma\mu\xi(\xi-1)} = \frac{\varphi_{\gamma p}}{\varphi_{\gamma T}} \,  . \label{eq34}
\end{gather}
The quantity $\varphi_{Tp}$ relates the relative change of temperature to the relative change of pressure depending on the value of $\gamma$. At $\gamma = 0$ we have $\varphi_{Tp} = \mu = R/c_p$ for an adiabatic process in dry air.

Note that the expression for vertical temperature gradient for adiabatic condensation $\Gamma \equiv -\pt T/\pt z \equiv T/h_T$ \eqref{eq28} can be obtained from \eqref{eq21} in the form of \eqref{eq34} only under the condition that the expression for condensation power \eqref{eq15} is known such that the continuity equation takes the form of \eqref{eq29}. If condensation power $\sigma$ in \eqref{eq8} remains unknown, then relations \eqref{eq29} do not exist, such that the derivatives $\pt p/\pt x$ and $\pt \gamma/\pt x$ in \eqref{eq21} remain unknown as well, leaving the relation between heights $h_T$ and $h_p$ in \eqref{eq34} undetermined.

In a horizontally bounded atmosphere where $\pt p/\pt x = 0$, $\pt \gamma/\pt x = 0$, see also \eqref{eq3}, from \eqref{eq8} we obtain that $\sigma = s_d$, i.e., at a given value of vertical velocity
$w$ setting the value of $\sigma$ will determine the quantities $\pt \gamma_d/\pt z$ \eqref{eq9} and $\pt \gamma/\pt z$ \eqref{eq8}. Then according to \eqref{sp} we have
$$\frac{1}{\widetilde{N}}\frac{\pt \widetilde{N}_v }{\pt z} = \frac{s_d}{w p}.$$
Moist air is not in hydrostatic equilibrium, $h_p \ne h_g$. Condensation causes an uncompensated upward force that accelerates the vertical air flow. Imposing the condition of hydrostatic equilibrium causes the vertical force to disappear and the process of condensation to discontinue. Then equations \eqref{eq32}-\eqref{eq34} assume the form
\beq\label{eq35}
\frac{1}{h_i}= \frac{1}{h_k}\,\Phi_{ik}\, , \quad (i \ne k)\,  ,
\eeq
where $i = \gamma, T$, $k = T,  p$, with functions $\Phi_{ik}$ defined in \eqref{eq24}-\eqref{eq26}. Owing to the fact that $h_p$ remains unspecified, expression \eqref{eq34} for the vertical temperature gradient $\Gamma$ is also undetermined in this case.

The replacement of the term $\gamma_d\xi$ in  \eqref{eq24}-\eqref{eq26}, \eqref{eq35}  by $\gamma (\xi-1)$ in \eqref{eq32}-\eqref{eq34} is related to the account of relation \eqref{eq29}, i.e., to the fact that the horizontal pressure gradient is not zero. However, owing to the large magnitude of $\xi$ ($\xi = 18$ at $T = 288$~Ê) and the small magnitude of $\gamma$, the values of $\varphi_{ik}$ in equations \eqref{eq32}-\eqref{eq34} differ from $\Phi_{ik}$ in \eqref{eq24}-\eqref{eq26} by a magnitude of about 2\%, which is beyond the accuracy of the existing measurements of $\Gamma$. Formulas \eqref{eq35}, \eqref{eq26} for $\Gamma \equiv T/h_T$ coincide with that found in the literature \cite{gill,hol04,eman} if one ignores the deviation of the moist air distribution from hydrostatic equilibrium and replaces $h_p$ for $h_g$.

In an atmosphere consisting of water vapor only we have $\gamma =1$, $\Phi_{Tp} = \xi^{-1}$, $M = M_v$. In this case after multiplying both parts of equation \eqref{eq35} by $\Phi_{Tp}^{-1}$ this equation and the last equality in \eqref{eq28} take the following form:
\beq\label{eq36}
\frac{\xi}{h_T} = \frac{1}{h_p} \, , \quad \frac{1}{h_\gamma}= \frac{\xi}{h_T} - \frac{1}{h_p} = 0\, .
\eeq
Then in hydrostatic equilibrium we have $h_p = h_v$, $\Gamma = T/\xi h_v = 1.2$~K/km \cite{hess07}; force $f_c$ \eqref{wfc} that ensures the condensation-induced ascent of moist air disappears, condensation stops and the atmosphere (neglecting the greenhouse effect) becomes vertically isothermal with $\Gamma = 0$.

\section{Conclusions}
\label{sec6}

In this work we have considered the physical principles of a large-scale circulation that arises due to water vapor condensation as the air ascends in the gravitational field of the Earth. In the absence of condensation the dynamic gas flows can arise in the gravitational field only at the expense of the Archimedes buoyancy linked to a horizontal temperature gradient \cite{land,eman}. Studies of water vapor condensation have been until very recently focused at the buoyancy effects associated with latent heat release \cite{acpd10,spe11}. The power of latent heat release upon condensation, which is $\xi$ times larger than the power of dynamic condensation-induced circulation, diminishes the rate of air cooling during its ascent in the terrestrial gravitational field. But it does not change either the amount of water vapor that undergoes condensation or
the condensation-induced average pressure drop in the atmospheric column or the decrease in the weight of the air column upon condensation. Latent heat release just distributes the condensation power over a larger height.

Adiabatic ascent of dry air at $\gamma = 0$ \eqref{eq34} leads to the well-known magnitude of the negative vertical lapse rate of air temperature equal to $\Gamma_d \equiv 9.8$ K/km. Condensation during adiabatic ascent of moist air leads to a release of latent heat which, in accordance to \eqref{eq34}, reduces the negative temperature lapse rate in the area where the ascent of moist air and condensation take place down to about $\Gamma_v \simeq 3$$-5$ K/km depending on the value of $\gamma$. In the area where the air undergoes descent (that is not accompanied by condensation), the lapse rate remains equal to $\Gamma_d$. Heat transfer, which is not related to circulation but which, because of circulation, has the form of intense turbulent processes, rapidly mixes the temperature lapse rates producing the mean tropospheric value of $\overline{\Gamma} \simeq (\Gamma_v + \Gamma_d)/2 =6.5$ K/km in the environment between the two areas. This gives rise to a positive buoyancy of the ascending air as compared to the ambient air, such that the Archimedes force arises that pushes the ascending air upwards; this force is proportional to the difference $\overline{\Gamma} - \Gamma_v \simeq 3.0$ K/km. Meanwhile in the area of descent the same air volume deprived of the condensed water vapor also experiences the Archimedes force that similarly pushes it upwards; this force is proportional to $\Gamma_d - \overline{\Gamma} \simeq  3.3$ K/km, i.e. it is approximately the same or larger than in the area of ascent. Summing the works that are opposite in sign -- as they are performed by upward-directed conservative forces acting on both the ascending and descending air volumes -- results in a significant decrease \cite{eman} (practical zeroing) of the cumulative potential energy related to buoyancy. This blocks the buoyancy-based circulation related to latent heat release, which therefore can represent but a minor correction to the considered condensation-induced circulation caused by the decrease of the number of molecules in the gas phase.


\begin{thebibliography}{21}
\bibitem[1]{gro64} G. M. Grover, T. P. Cotter, and G. F. Erickson, \newblock J. Appl. Phys. {\bf 35}, 1990 (1964).
\bibitem[2]{xie08} X. L. Xie, Y. L. He, W. Q. Tao, and H. W. Yang, \newblock  Appl. Therm. Eng. {\bf 28}, 433 (2008).
\bibitem[3]{sha90} P. N. Shankar and M. D. Deshpande, \newblock  Phys. Fluids A {\bf 2}, 1030 (1990).
\bibitem[4]{son00} Y. Sone, \newblock Transp. Theory Stat. Phys. {\bf 29}, 227 (2000).
\bibitem[5]{ryk09} V. A. Rykov, V. A. Titarev, and E. M. Shakhov, \newblock Fluid Dynamics {\bf 44}, 464 (2009).
\bibitem[6]{pl11a} A. M. Makarieva and V. G. Gorshkov,  \newblock  Phys. Lett. A {\bf 375}, 1053 (2011).
\bibitem[7]{pl11b} A. M. Makarieva, V. G. Gorshkov, and A. V. Nefiodov, \newblock  Phys. Lett. A {\bf 375}, 2259 (2011).
\bibitem[8]{gill} A. E. Gill, \newblock {\it Atmosphere-Ocean Dynamics}, Acad. Press, New York (1982).
\bibitem[9]{smi97} R. K. Smith, \newblock Q. J. R. Meteorol. Soc. {\bf 123}, 407 (1997).
\bibitem[10]{hess07} A. M. Makarieva and V. G. Gorshkov, \newblock Hydrol. Earth Syst. Sci. {\bf 11}, 1013 (2007).
\bibitem[11]{bry09} G. H. Bryan and R. Rotunno, \newblock Month. Weather Rev. {\bf 137}, 1770 (2009).
\bibitem[12]{pel11} J. Pelkowski and T. Frisius, \newblock J. Atmos. Sci. {\bf 68}, 2430 (2011).
\bibitem[13]{land} L. D. Landau and E. M. Lifshitz, \newblock {\it Theoretical Physics. Fluid Mechanics}, GITTL, Moscow (1954).
\bibitem[14]{pl09a} A. M. Makarieva and V. G. Gorshkov, \newblock Phys. Lett. A {\bf 373}, 2801 (2009).
\bibitem[15]{wat10} A. M. Makarieva and V. G. Gorshkov, \newblock Int. J. Water {\bf 5}, 365 (2010).
\bibitem[16]{pl09b} A. M. Makarieva and V. G. Gorshkov, \newblock Phys. Lett. A {\bf 373}, 4201 (2009).
\bibitem[17]{hol04} J. R. Holton, \newblock {\it An Introduction to Dynamic Meteorology}, Acad. Press, Amsterdam (2004).
\bibitem[18]{lev} V. G. Levich, \newblock {\it Course in theoretical physics, V. 1}, Fizmatlit, Moscow (1962).
\bibitem[19]{eman} K. A. Emanuel, \newblock {\it Atmospheric convection}, Oxford Univ. Press, Oxford (1994).
\bibitem[20]{acpd10} A. M. Makarieva, V. G. Gorshkov, D. Sheil, A. D. Nobre, and B.-L. Li, \newblock  Atmos. Chem. Phys. Discuss.  {\bf  10}, 24015 (2010).
\bibitem[21]{spe11} T. Spengler, J. Egger, and S. T. Garner, \newblock J. Atmos. Sci. {\bf 68}, 347 (2011).
\end{thebibliography}
\end{document}